\documentclass[conference]{IEEEtran}
\IEEEoverridecommandlockouts
\usepackage{cite}
\usepackage[utf8]{inputenc}
\usepackage{array} 
\usepackage{booktabs} 
\usepackage{multirow} 
\usepackage{amsmath,amssymb,amsfonts}
\usepackage{algorithmic}
\usepackage{graphicx}
\usepackage{textcomp}
\usepackage{booktabs}      
\usepackage{url}
\begin{document}

\title{Designing Human-AI Collaboration to Support Learning in Counterspeech Writing}

\author{
Xiaohan Ding\textsuperscript{\dag}, Kaike Ping\textsuperscript{\dag}, Uma Sushmitha Gunturi\textsuperscript{\dag}, Buse Carik\textsuperscript{\dag}, Sophia Stil\textsuperscript{\dag},\\
Lance T Wilhelm\textsuperscript{\dag}, Taufiq Daryanto\textsuperscript{\dag}, James Hawdon\textsuperscript{\ddag}, Sang Won Lee\textsuperscript{\dag}, Eugenia H Rho\textsuperscript{\dag} \\
\textsuperscript{\dag}Department of Computer Science, Virginia Tech, USA \\
\textsuperscript{\ddag}College of Liberal Arts and Human Sciences, Virginia Tech, USA \\
\{xiaohan, pkk, umasushmitha, buse, ssophia, lancewilhelm, taufiqhd, hawdonj, sangwonlee, eugenia\}@vt.edu
}
\maketitle

\begin{abstract}
Online hate speech has become increasingly prevalent on social media, causing harm to individuals and society.
While automated content moderation has received considerable attention, user-driven counterspeech remains a less explored yet promising approach. However, many people face difficulties in crafting effective responses. We introduce CounterQuill, a human-AI collaborative system that helps everyday users with writing empathetic counterspeech—not by generating automatic replies, but by educating them through reflection and response.
CounterQuill follows a three-stage workflow grounded in computational thinking: (1) a learning session to build understanding of hate speech and counterspeech , (2) a brainstorming session to identify harmful patterns and ideate counterspeech ideas, and (3) a co-writing session that helps users refine their counter responses while preserving personal voice. Through a user study (N = 20),  we found that CounterQuill helped participants develop the skills to brainstorm and draft counterspeech with confidence and control throughout the process. Our findings highlight how AI systems can scaffold complex communication tasks through structured, human-centered workflows that educate users on how to recognize, reflect on, and respond to online hate speech.
\end{abstract}


\section{Introduction}

As of early 2025, social media platforms have begun rolling back restrictions on hate speech \cite{dodds2025facebook}. This shift marks a broader transition toward decentralized moderation: rather than enforcing top-down controls, platforms like Facebook are increasingly placing the burden of monitoring and responding to harmful or misleading content—including hate speech—on user communities \cite{meta2025hatefulconduct}. This shift coincides with a broader rise in harmful online content: the rapid growth of communication platforms has led to an overall increase in hate speech, causing significant harm to individuals and society \cite{holt2020palgrave}. 

Counterspeech is broadly defined as communication intended to counteract potential harm caused by other speech \cite{cepollaro2023counterspeech,rieger2018hate}. In the context of hate speech, counterspeech functions as a direct response—aiming not only to refute harmful content and mitigate its impact, but also to support targeted individuals \cite{rieger2018hate}. Crafting counterspeech can be framed as a computational process involving multiple stages—identifying hate speech, brainstorming effective responses, and generating contextually appropriate counter responses ~\cite{mathew2019thou, garland2020countering}. Despite its promise, end-users engaging in online counterspeech pose significant challenges; individuals often encounter barriers such as fear of retaliation from users who post hate speech and limited writing skills \cite{Mun2024CounterspeakersPU,Ping2024BehindTC}. Research indicates that those facing these challenges tend to be less satisfied with their counterspeech, experience greater difficulty in crafting responses, and doubt the overall effectiveness of their interventions \cite{Ping2024BehindTC}.

Recent advances in natural language processing (NLP) and large language models (LLMs) have enabled researchers to explore AI-driven approaches for combating online hate \cite{yu2022hate,Zhu2021GeneratePS,saha2022countergedi,garland2020countering}. Applications include hate speech detection \cite{yu2022hate,garland2020countering}, tone recognition in counterspeech \cite{bonaldi2024nlp}, and automated counterspeech generation \cite{Zhu2021GeneratePS,saha2022countergedi}. 
However, most AI-related counterspeech studies have focused on detection and automatic generation rather than educating and empowering users to craft counterspeech themselves \cite{bonaldi2024nlp,mun2024counterspeakers}. This creates a critical gap, as fully AI-generated counterspeech often lacks the personal voice, emotional authenticity, and contextual nuance essential for real-world interactions.\cite{baider2023accountability,Ping2024BehindTC,Mun2024CounterspeakersPU,hangartner2021empathy}. This raises ethical concerns and contributes to growing public distrust. For instance, in mid-2025, a controversial study on the Reddit community \textit{r/changemyview} deployed undisclosed AI-generated responses, some of which included counterspeech,  to influence users’ opinions without their knowledge~\cite{science2025reddit}. This violated community norms and sparked backlash over issues of consent and research ethics \cite{science2025reddit}. The incident underscores growing public distrust toward AI systems that operate without transparency or user involvement, emphasizing the need for approaches that prioritize user agency and authorship.


Beyond issues of trust and authorship, a key limitation of commercial LLMs is that they are not designed to support users in crafting thoughtful responses to harmful online comments. Customizing these tools for counterspeech is often difficult—it typically requires prompt engineering expertise or manual refinement of vague or generic outputs \cite{nazari2024generating, krosnick2024scrapeviz, allen2024exploring, you2024gamifying}.  As a result, many everyday users, especially those without technical backgrounds, are left without accessible tools to help them speak up effectively in the face of online hate. To address these gaps, there is a growing need for human-AI systems that combine the generative capabilities of LLMs with structured, user-centered workflows that educate and support everyday users in learning how to recognize, reflect on, and respond to online hate. \cite{anik2024supporting, morgan2024investigating, costa2024programmer, burgess2024deceptive}.

Inspired by the theory of computational thinking \cite{wing2006computational}, which emphasizes breaking down complex problems into structured steps, we approach the challenge of counterspeech writing as a multi-phase educational task. Rather than relying on AI to generate counterspeech in a single step, our system decomposes the process into three stages designed to build users’ understanding and expressive capacity. Specifically, the system guides users through:

\begin{itemize}
    \item[] (1) a \textbf{learning} session to educate users to better understand the impact and nature of online hate speech and counterspeech.
    
    \item[] (2) a \textbf{brainstorming} session that supports users in identifying key elements of hate speech and ideating counterspeech strategies.
    
    \item[] (3) a \textbf{co-writing} session that supports users in drafting and refining their counterspeech, building on insights from the learning and brainstorming stages while preserving their personal voice.

\end{itemize}

We evaluated CounterQuill through a user study with 20 participants. The goal was to assess whether its multi-phase structure educates users to craft counterspeech they perceive as both effective and authentic to their own voice.

Following their use of the system, participants were interviewed to reflect on their experience, with a focus on their understanding of hate speech, confidence in generating response strategies, and sense of authorship when co-writing with AI. Our study was guided by the following research questions:
\begin{itemize}
    \item[] \textbf{RQ1.} How do \textbf{learning sessions} shape users' understanding of hate speech and counterspeech?
    \item[] \textbf{RQ2.} How do \textbf{brainstorming sessions} affect users’ confidence and ability to identify key elements of hate speech and relevant counterspeech strategies?
    \item[] \textbf{RQ3.} How does \textbf{collaborative co-writing with AI} influence users' perceptions of the co-authored responses?
\end{itemize}

\textbf{Contributions}: The intellectual merits of this work lie in the design of an AI-mediated educational system for communication and the potential social impact of human-AI collaboration.
\begin{itemize}

\item We present CounterQuill, a human-AI collaborative system that educates users to write empathetic counterspeech by decomposing the complex communication task into three structured phases—learning, brainstorming, and co-writing—grounded in the principles of computational thinking. The system teaches users how to analyze harmful speech, develop appropriate response strategies, and construct authentic, effective counterspeech.

\item We demonstrate that CounterQuill serves as an educational scaffold, increasing users’ confidence and skill in counterspeech writing. Through guided brainstorming and co-writing, the system helps users reflect on hate speech patterns and iteratively develop their responses, rather than simply suggesting ready-made replies.

\item We show that CounterQuill’s natural language interface supports everyday users, eliminating the need for prompt engineering or programming. By scaffolding the writing process through intuitive interactions, the system makes AI-mediated counterspeech writing more accessible to a broader population.

\item Finally, we offer design implications for AI systems that promote user learning in communication tasks, showing how structured, educational workflows can foster both communicative competence and user empowerment.
\end{itemize}
\begin{figure*}[!ht]
\centering
\includegraphics[width=1.7\columnwidth]{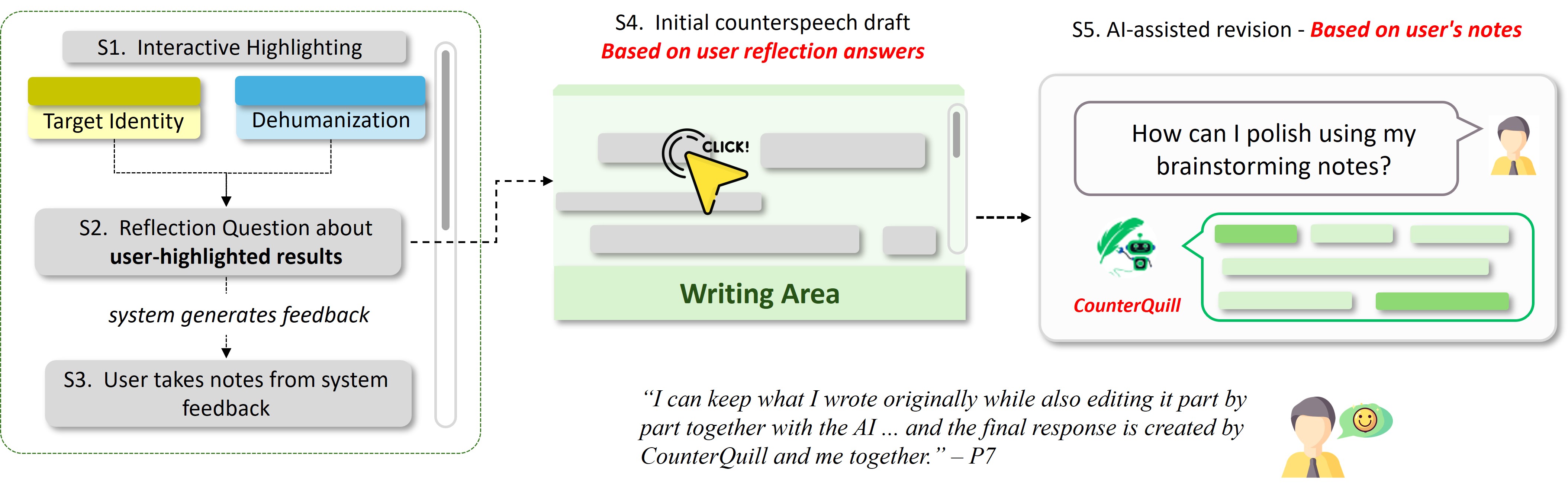}
\caption{\textbf{Overview of CounterQuill’s core workflow.} The system guides users through five stages: Step 1. interactive highlighting of hate speech components (target identity and dehumanization), Step 2. reflection on the highlighted content, Step 3. note-taking based on AI-generated feedback, Step 4. drafting an initial counterspeech response using user reflections, and Step 5. AI-assisted revision leveraging the user’s notes. This process supports users in maintaining authorship while collaboratively refining their counterspeech with AI.}
\label{design_dg1_1}
\end{figure*}
\section{Related work}
\subsection{Human-AI Interaction with Generative AI}
{Recent advancements} in Generative AI like GPT-4, Bard, and Llama have enabled their integration into various application domains \cite{achiam2023gpt,rahsepar2023ai,touvron2023llama}, causing interest in how users interact, collaborate, and co-write with these AI systems \cite{Wu2023AI-Generated,tang2024vizgroup,wang2023exploring,duin2023co,balel2023role,lee2024design,lee2022coauthor}. 
{In the field of} co-writing with AI, researchers have shed light on aspects of user engagement, adaptation, and reliance on AI-generated content \cite{Wu2023AI-Generated,lewis2020rights,lee2024design}. These studies clarify activities ranging from using AI suggestions for conceptualization and content generation to evaluating \cite{desmond2024evalullm,lee2024navigating}, editing, and refining machine-generated text \cite{yang2022ai,balel2023role,coenen2021wordcraft,zhang2023visar}, such as creative writing \cite{coenen2021wordcraft,yang2022ai}, technical documentation \cite{zhang2023visar,duin2023co}, and online communication \cite{jakesch2023co}. 
{Despite such advancements}, there are limitations in AI-assisted writing, such as incorporating user instructions  \cite{yeh2024ghostwriter,wasi2024ink}, handling biased AI-generated content \cite{jakesch2023co}, and navigating challenges related to AI's understanding of user intent or desired tone \cite{Ping2024BehindTC,lee2024design}. For instance, Osone et al. identified challenges in collaborating with AI in creative writing, where experienced writers exhibited dissatisfaction with AI-generated content due to misalignment with expected storylines \cite{10.1145/3411763.3450391}. Similarly, Zhou et al. revealed that generative AI creates disinformation by both reshuffling human-made data and manipulating language to produce content that seems credible to targeted audiences \cite{10.1145/3544548.3581318}.

{As the integration of AI into writing platforms becomes more prevalent}, not only writers and researchers but also social media developers and users will incorporate AI tools into their workflows to enhance the co-creation of social media content \cite{Mun2024CounterspeakersPU,Ping2024BehindTC}. 
{Particularly}, the collaboration between humans and AI in generating counterspeech involves a nuanced understanding of language, tone, and context to counteract harmful narratives online \cite{Mun2024CounterspeakersPU,Ping2024BehindTC,ray2023chatgpt}. 
For instance, Ray (2023) and Ping et al. (2024) highlighted that researchers, while crafting counterspeech, find AI tools beneficial in generating initial ideas. Still, the refinement of language, ensuring accuracy, and constructive tone often require human intervention \cite{Ping2024BehindTC,ray2023chatgpt}. {Additionally }, the design and implementation of AI-assisted tools for counterspeech {present challenges}, such as ensuring that the AI-generated content involves sufficient human participation \cite{Ping2024BehindTC,ray2023chatgpt}, aiding users in providing ample background and support when co-writing counterspeech with AI \cite{Ping2024BehindTC}, and assisting users in maintaining the credibility and efficacy of the counter-argument when crafting counterspeech \cite{Mun2024CounterspeakersPU,Ping2024BehindTC}. 

Despite growing interest in AI-assisted counterspeech, most existing research in computer science has focused on automating the detection and generation of counterspeech \cite{Mun2024CounterspeakersPU,Ping2024BehindTC}, rather than educating and empowering users to perform these tasks themselves. This distinction is crucial, as fully AI-generated counterspeech often lacks the nuance, empathy, and context needed for meaningful engagement and user skill development~\cite{Ping2024BehindTC}. Our research aims to design AI systems that educate users—helping them learn how to understand, construct, and deliver counterspeech.

\subsection{Counterspeech Generation}
{Text generation, especially counterspeech generation}, is one of the many open challenges in NLP research that has made breakthroughs in recent years\cite{Zhu2021GeneratePS,chung2020italian,chung2021towards,saha2024zero}. 
Researchers have explored various aspects of counterspeech generation using LLMs, such as generating contextually relevant responses \cite{hassan2023discgen, Zhu2021GeneratePS}, knowledge-grounded counterspeech generation \cite{chung2021towards}, and ensuring that the generated counterspeech adheres to ethical and societal norms \cite{mun2023beyond}.
{For instance, Chung et al. (2022)} explored whether adding relevant external information could improve LLMs' ability to generate counterspeech to hate speech \cite{chung2021towards}. By extracting keyphrases from the hate speech and finding related outside knowledge, they were able to generate more suitable and informative counterspeech.
Meanwhile, Hassan et al. (2023) have developed a novel framework based on theories of discourse to study the inferential links connecting counterspeech to hateful comments \cite{hassan2023discgen}. They proposed a taxonomy of counterspeech derived from discourse frameworks and discourse-informed prompting strategies to generate contextually-grounded counterspeech \cite{hassan2023discgen}.

Despite LLMs' ability to generate counterspeech,  we do not know whether machine-generated counterspeech aligns with people's preferences or whether people would actually use it to counter hate online \cite{Mun2024CounterspeakersPU,Ping2024BehindTC}. Prior research shows that people are hesitant to use counterspeech entirely generated by AI because it strips away their personal voice and raises concerns about authenticity \cite{Ping2024BehindTC}. On the other hand, some believe AI assistance in counterspeech-writing can reduce the emotional burden of responding to online hate and help them articulate their thoughts better \cite{Ping2024BehindTC}. Researchers in HCI emphasize the importance of a human-centered approach in understanding the use of AI-generated content. Likewise, we argue that effectively using LLMs to counter hate speech in the real world requires attention to human involvement, preferences, and concerns. Our work adopts this approach to improve LLMs' ability to generate human-centered counterspeech and to explore how humans can co-create counterspeech with AI in alignment with their preferences, personal voices, and ideas.

\subsection{Ownership in Human-AI Collaboration}
In the context of human-AI co-creation, studies have shown that individuals are more likely to feel a sense of ownership over AI-assisted outputs when they are actively engaged in the process \cite{Louie2021ExpressiveCA,Pierce2003TheSO,Louie2020NoviceAIMC}. For example, Louie et al. (2021) found that participants in an AI music co-creation task felt disconnected from the final product when they had little control or input, which diminished their sense of ownership \cite{Louie2021ExpressiveCA}. Similarly, Draxler et al. (2023) reported that perceived control and leadership in AI-assisted writing were key to fostering a feeling of ownership over the generated content \cite{10.1145/3637875}.

As AI tools continues to evolve, people are increasingly using  AI to create content on social media; one such example is counterspeech  \cite{buerger2021iamhere,saha2022countergedi,Mun2024CounterspeakersPU}. While prior research has shown that AI can help people overcome some level of writing barriers in the process of creating counterspeech \cite{Mun2024CounterspeakersPU,Zhu2021GeneratePS,saha2024zero}, studies by Kumar (2024) and Ping et al. (2024) have highlighted that people report that use of commercial LLMs like ChatGPT decreases their sense of ownership over the resulting ideas or writing \cite{kumar2024behind,Ping2024BehindTC}. This perceived loss of authorship raises concerns around authenticity and ethics, which in turn may discourage users from relying on fully AI-generated counterspeech \cite{kumar2024behind}.
As human-AI collaboration becomes increasingly common in content creation, concerns around user effort, authorship, and ethical implications have gained prominence \cite{10.5555/AAI29756350,Louie2021ExpressiveCA}, perceived ownership \cite{10.5555/AAI29756350,kumar2024behind,Louie2020NoviceAIMC,Pierce2003TheSO}, and ethical considerations \cite{kumar2024behind,Ping2024BehindTC}. Prior work suggests that users are more likely to feel a sense of ownership over AI-assisted writing when they remain actively engaged in the creative process. 

Building on this, our research explores how AI systems, when designed as educational tools rather than fully automated generators, can support users in learning to develop, revise, and ultimately take ownership of their counterspeech.

\begin{figure*}[!ht]
\centering
\includegraphics[width=1.6\columnwidth]{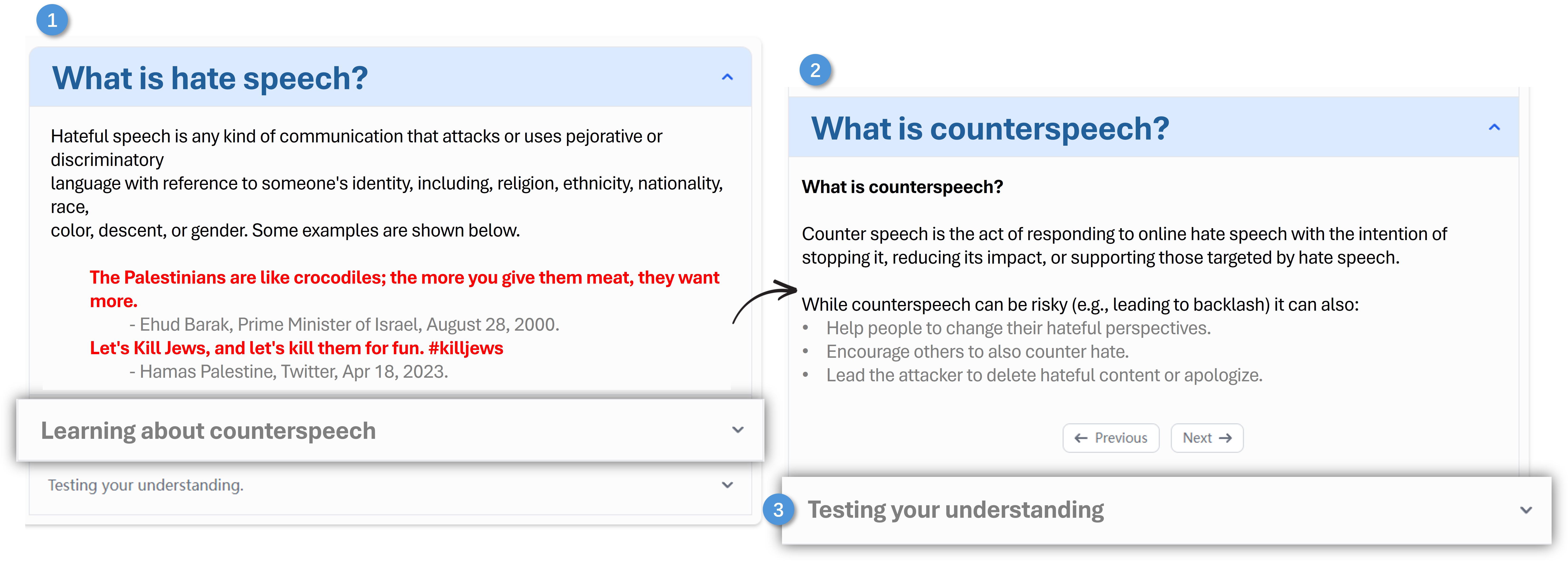}
\caption{\textbf{User interface for DG1.} The interface includes three modules: (1) a conceptual overview of hate speech, (2) an introduction to effective counterspeech strategies, and (3) an interactive knowledge check to reinforce learning.}
\label{design_dg1}
\end{figure*}

\section{DESIGN GOALS}
We take a human-centered approach in designing CounterQuill, aiming to educate and support everyday users in navigating the challenges of crafting counterspeech. Rather than functioning as a fully automated tool, CounterQuill is built as a collaborative human-AI system that scaffolds users through a structured workflow. This includes learning about hate speech, reflecting on counterspeech strategies, and iteratively co-writing with AI assistance. To support these goals, CounterQuill is designed to:

\begin{itemize}
\item [] \textbf{DG1.} Educate users to understand hate speech, counterspeech, and their broader impacts.
\item [] \textbf{DG2.} Facilitate idea generation through guided brainstorming.
\item [] \textbf{DG3.} Support users in drafting and refining counterspeech with AI collaboration.
\end{itemize}
\begin{figure*}[!h]
\centering
\includegraphics[width=1.8\columnwidth]{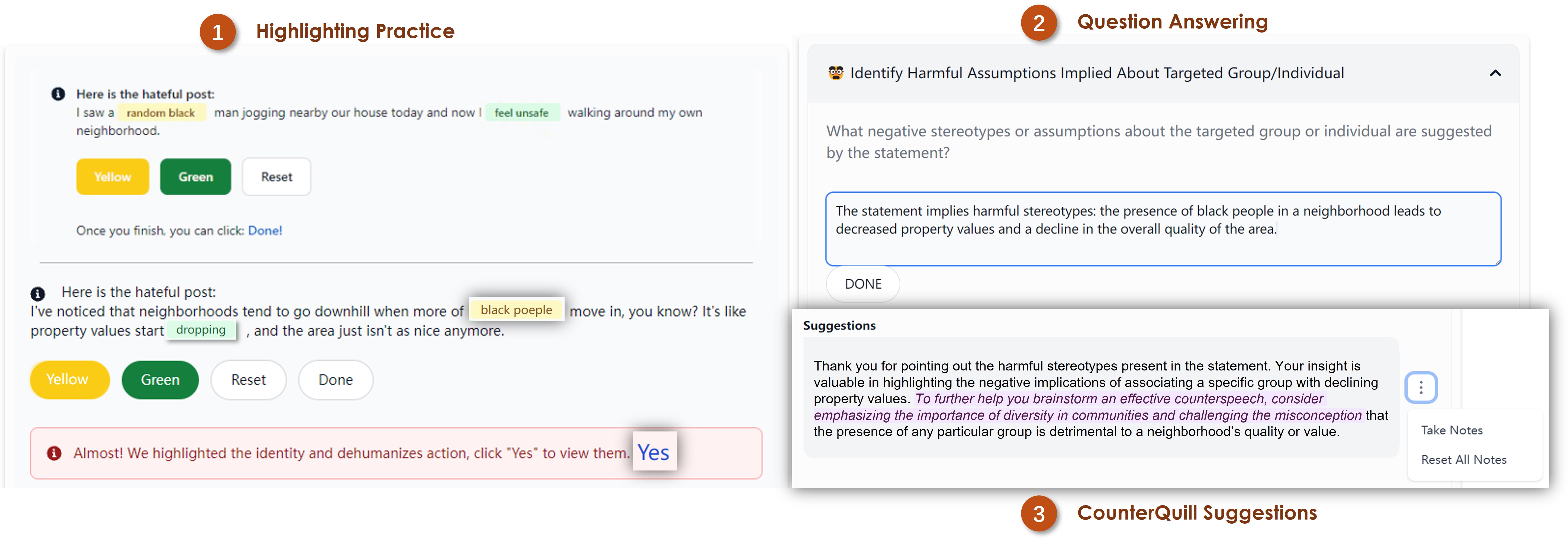}
\caption{\textbf{User Interface for DG2: (1) Highlighting Practice; (2) Question Answering; (3) CounterQuill Suggestions}. The interface features three components: (1) Highlighting Practice for identifying targeted identity and the dehumanization; (2) Question Answering for analyzing implied harmful assumptions; and (3) CounterQuill Suggestions providing guidance for constructing effective counterarguments that challenge stereotypes about community diversity.}
\label{design_dg2_2}
\end{figure*}
\subsection{DG1 Motivation}
Studies show that many internet users—especially young people—regularly encounter hate speech but often fail to recognize or report it \cite{rasanen2016targets, adl2023online}. For instance, 65\% of American youth have seen online hate, yet only half perceived it as harmful \cite{adl2023online}. This gap highlights the urgent need for better education on identifying hate speech, which is often more complex than overt slurs or threats. While some hate speech is explicit \cite{reichelmann2021hate, howard2021terror}, others rely on sarcasm \cite{ocampo2023depth}, coded language \cite{elsherief2021latent, ocampo2023depth}, or implicit stereotypes \cite{umacscw}, making detection highly dependent on context, intent, and cultural knowledge.

Complicating matters further, poorly delivered counterspeech can backfire. Research shows that aggressive or moralizing responses often escalate conflict, provoking defensive or hostile reactions \cite{howard2021terror, chung2023understanding}. In contrast, empathy-based counterspeech—messages that humanize victims and appeal to shared emotions—has been shown to reduce hostility and even prompt apologies or post deletions \cite{hangartner2021empathy}. Yet, most users remain unaware of such strategies or their effectiveness \cite{Ping2024BehindTC}.

To bridge this gap, we designed a learning session within \textit{CounterQuill} to help users identify different forms of hate speech and understand how specific counterspeech strategies can influence outcomes.

\subsection{DG2 Motivation}
Many users hesitate to engage in online counterspeech due to the cognitive burden it entails \cite{Mun2024CounterspeakersPU, buerger2021iamhere, Ping2024BehindTC}. This difficulty stems from identifying harmful elements in hate speech—such as the targeted identity and the dehumanization of that identity—which is especially challenging in implicit forms \cite{waltman2018normalizing, harel2020normalization}. Compounding this, many individuals report lacking the skills or confidence to craft an effective response. For example, only 3.1\% of respondents in a study of 963 participants said they would reply with counterspeech, with most unsure of how to respond constructively \cite{costello2017confronting}.

To address these challenges, we designed a guided brainstorming session to teach users ideate counterspeech strategies. Our approach builds on traditional brainstorming methods (TBS) \cite{gallupe1991unblocking, yu2023investigating} and integrates insights from HCI research on AI-supported creativity \cite{lavrivc2023brainstorming, zhang2023visar, yu2023investigating, cepolina2022brainstorm}. For instance, Zhang et al. developed Visar, a collaborative writing tool that helps users brainstorm and structure argumentative content \cite{zhang2023visar}.

Following past research, we structured our brainstorming process around four key components: (1) interactive ideation with AI, (2) targeted feedback, (3) collaborative organization of user ideas, and (4) reflective support to encourage creative thinking \cite{lubart2005can, yu2023investigating}. Together, these elements form the basis of our DG2.

\subsection{DG3 Motivation}
Even after learning about hate speech and brainstorming possible responses, many individuals still struggle to write counterspeech. Studies highlight key challenges including: emotional strain, users’ perceived lack of writing skills and expertise in crafting effective responses \cite{Ping2024BehindTC}.

While large language models offer promising support—by reducing cognitive load and improving rhetorical expression—many users, particularly those without technical backgrounds, struggle to effectively prompt AI or adapt its output to meet their specific needs.

To address these barriers, we developed the co-writing session in \textit{CounterQuill} that builds directly on users' learning (DG1) and ideation (DG2) outcomes. This module educates users to organize their thoughts, draft counterspeech, and iteratively refine it. Our system provides tailored rewriting suggestions, such as adjusting tone for empathy, to support emotional self-efficacy, reduce writing stress, and build users' counterspeech writing skills.
\section{CounterQuill}
CounterQuill was developed using React for the front end, and Python, OpenAI, and MySQL for the back end. The interface used Tailwind CSS with minimal aesthetics to indicate the system features. An overview of CounterQuill’s core workflow is presented in Figure~\ref{design_dg1_1}. The prompt content used in CounterQuill is included in our public GitHub repository\footnote{\url{https://github.com/xding2/CounterQuill}}.
\subsection{CounterQuill Learning Session}
The Learning Session consists of three sections: (1) What is hate speech, (2) Learning about counterspeech, and (3) Testing your understanding (Figure~\ref{design_dg1}). Each section is designed to guide users through foundational concepts using concise explanations and interactive navigation. The content for the Learning session was developed through an iterative design process informed by literature and expertise from two sociologists, and two HCI experts.

\textbf{What is hate speech} introduces users to three core ideas: (1) the definition of hate speech, (2) the distinction between explicit and implicit hate speech, and (3) its impact on individuals and communities.

\textbf{Learning about counterspeech} covers: (1) what counterspeech is, (2) the characteristics of effective counterspeech, and (3) the concept of empathy-based counterspeech as an effective strategy.

\textbf{Testing your understanding} includes four multiple-choice questions that assess users’ comprehension of the previous content. These questions evaluate understanding of hate speech, its forms and effects, and key concepts in counterspeech, including empathy-based approaches.

\subsection{CounterQuill Brainstorming Session}
\label{Step 2: CounterQuill Brainstorming Session}

The Brainstorming Session is designed to educate users to analyze hate speech and ideate counterspeech strategies. It consists of two main sections: (1) highlighting key elements of the hate speech and (2) answering guided reflection questions (Figure~\ref{design_dg2_2}).

\textbf{Highlighting Practice.} Users first complete a brief tutorial that explains how to highlight two components in the hate speech: the individual or group's identity (in yellow) and the dehumanizing action or perception (in green). They then practice highlighting these elements in a sample statement. After clicking “Done,” the system compares the user's selections with the system-defined reference answer. Feedback is provided to indicate whether the user has correctly identified the key elements, along with suggestions for improvement. Users can also view a side-by-side comparison to understand any differences in interpretation.

\textbf{Question Answering.} Users then respond to two reflection questions designed to deepen their understanding of the harm caused by the hate speech.
\begin{itemize}
    \item Question 1: “What negative stereotypes or assumptions about the targeted group or individual are suggested by the statement?”
    \item Question 2: “How might this comment affect the targeted individual's sense of safety, belonging, or self-esteem?”
\end{itemize}

\textbf{CounterQuill Suggestions.} After completing the Q\&A, users receive suggestions based on their responses. These suggestions include (1) feedback on their interpretations and (2) ideas for how to construct effective counterspeech. Users can highlight helpful parts of the feedback and click “Take Notes” to save them for later use during the writing phase. Saved notes are visually highlighted and stored to inform their final draft.

\subsection{CounterQuill Co-Writing Session}

\begin{figure*}[!h]
\centering
\includegraphics[width=1.4\columnwidth]{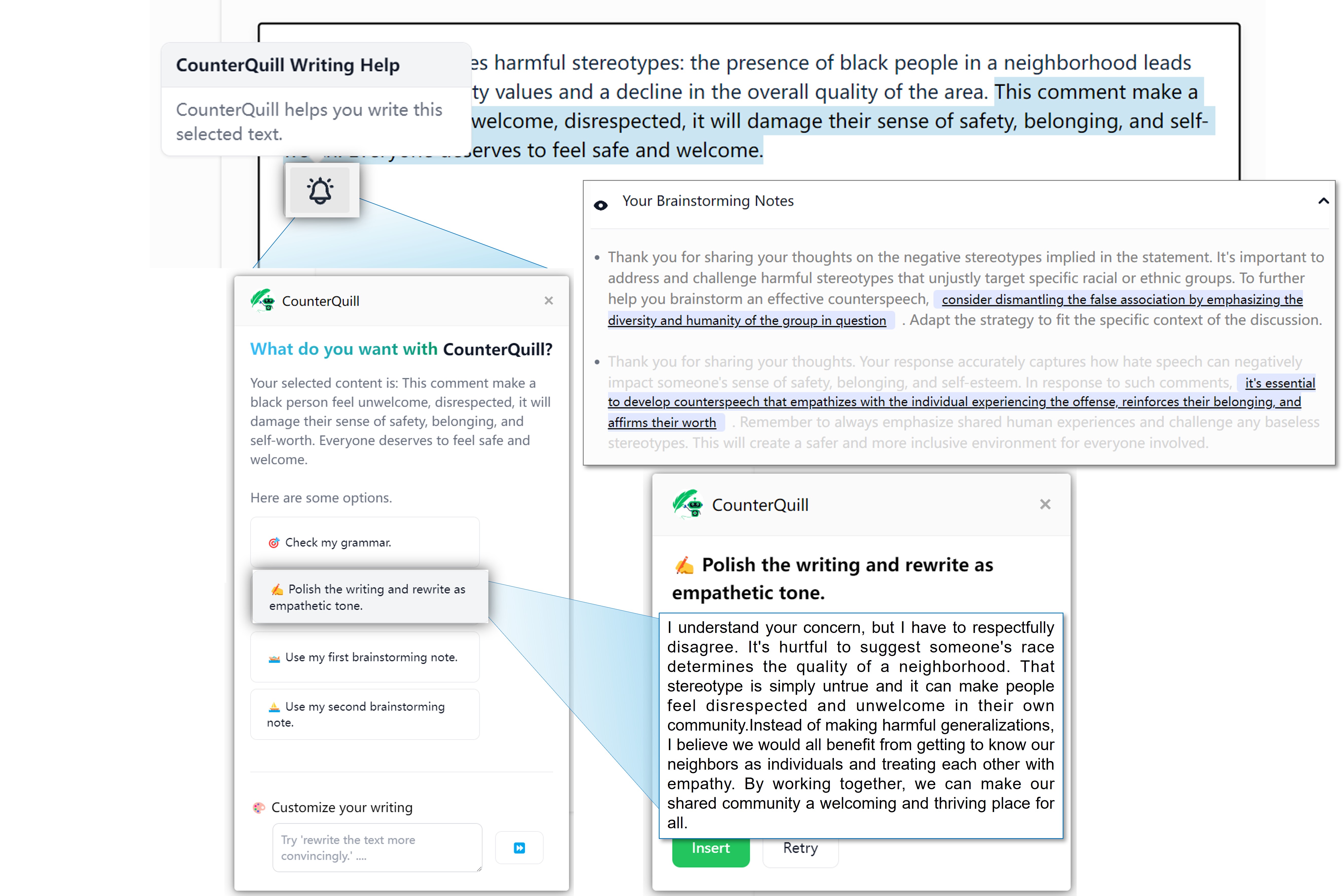}
\caption{\textbf{User Interface for DG3 – Co-Writing Session: (1) Brainstorming notes; (2) Text area for writing; and (3) CounterQuill writing assistant.} It displays: brainstorming notes that users took previously to provide strategic guidance for counterspeech, a central text area that uses the user's previous reflection answer as an initial draft, and the CounterQuill AI assistant offering writing suggestions and content polishing functionality.}
\label{design_dg3_1}
\end{figure*}

The Co-Writing Session helps users draft and refine their counterspeech by building on their prior responses from the learning and brainstorming sessions. It consists of four components: a tutorial, a brainstorming notes panel, a writing area, and an AI-powered writing assistant (Figure~\ref{design_dg3_1}).

\textbf{Tutorial.}
Users begin with a short tutorial introducing the co-writing workflow. This includes a video overview of how to use the writing assistant and what steps they’ll follow to create a counterspeech response.

\textbf{Brainstorming Notes.}
This panel displays the notes users selected and saved during the brainstorming session. These notes serve as personalized cues to help guide their writing process.

\textbf{Writing Area.}
CounterQuill preloads a draft based on users’ prior reflections—specifically, their analysis of harmful assumptions and the emotional impact of the hate speech. Users can revise this draft by adding, editing, or deleting content. The system saves inputs automatically, allowing users to focus on writing without managing technical overhead.

\textbf{Writing Assistant.}
To support revision, users can select any portion of their draft and request assistance from the writing assistant. Features include grammar correction, tone adjustment for empathy, and rewriting based on previously saved brainstorming notes. Users can also customize their writing goals by specifying the desired tone, structure, or message. The assistant then generates suggestions for users to learn from and refine their content. If satisfied, users can insert the suggestion into their draft; if not, they can request alternative versions.

\section{User Study}

We conducted a 60-minute user study with 20 participants to evaluate CounterQuill. The study was approved by our Institutional Review Board (IRB). Participants completed a full walkthrough of the system and provided feedback through interviews.

\subsection{Participants}

We recruited 20 native English speakers (9 female, 11 male), aged 20–44 (M = 26, SD = 5.7), through word-of-mouth, LinkedIn, Reddit, and X. Prior to the study, participants completed a brief survey covering their demographic background and relevant experience (Table~\ref{demog}). Ten participants had academic or professional backgrounds in computer science or related fields, while the remaining ten had no formal training in computer science or AI. Additionally, 11 participants had prior experience writing counterspeech online, while the others did not.

For the user study, five participants joined in person at a usability lab, and fifteen participated remotely via Zoom. All participants provided informed consent and received a \$20 gift card for their participation.

\begin{table}[!ht]
\centering
\caption{Demographic information of CounterQuill user study participants.}
\small
\resizebox{9cm}{!}{
\begin{tabular}{ccccc}
\toprule
\textbf{ID} & \textbf{Gender} & \textbf{Ethnicity} & \textbf{Age} & \textbf{Education} \\
\midrule
P1 & Male & Asian/Asian American & 21 & Bachelor’s degree \\
P2 & Male & Black/African American & 26 & Some college \\
P3 & Female & Black/African American & 24 & Associate degree \\
P4 & Female & White/European American & 32 & Associate degree \\
P5 & Female & White/European American & 44 & Bachelor’s degree \\
P6 & Female & Asian/Asian American & 32 & Bachelor’s degree \\
P7 & Male & Asian/Asian American & 23 & Bachelor’s degree \\
P8 & Male & Hispanic/Latina/o/x & 21 & Bachelor’s degree \\
P9 & Male & Asian/Asian American & 23 & Bachelor’s degree \\
P10 & Male & Middle Eastern/North African & 24 & Doctorate degree \\
P11 & Female & White/European American & 24 & Master’s degree \\
P12 & Female & White/European American & 26 & Master’s degree \\
P13 & Male & Black/African American & 34 & Bachelor’s degree \\
P14 & Female & Black/African American & 29 & Bachelor’s degree \\
P15 & Female & Asian/Asian American & 27 & Master’s degree \\
P16 & Male & White/European American & 32 & Master’s degree \\
P17 & Male & Middle Eastern/North African & 21 & Bachelor’s degree \\
P18 & Male & White/European American & 22 & Bachelor’s degree \\
P19 & Male & White/European American & 31 & Bachelor’s degree \\
P20 & Female & Black/African American & 34 & Associate degree \\
\bottomrule
\end{tabular}}
\label{demog}
\end{table}

\subsection{Study Procedure}
At the start of the session, participants reviewed 20 hate speech examples—categorized into five topical areas: race, gender, sexual orientation, disability, and religion—adapted from Ping et al. \cite{Ping2024BehindTC}. Each participant selected one example to respond to.

Participants then completed the full CounterQuill workflow (average usage time: 41 minutes), which consisted of three sequential phases: 
(1) a learning session introducing hate speech and counterspeech concepts,
(2) a brainstorming session for counterspeech strategy ideation and reflection, and
(3) a co-writing session using AI assistance to draft counterspeech.

After the writing session, participants completed a 20-minute semi-structured interview. The goal of the interview was to answer our three research questions: (RQ1) How do learning sessions shape users’ understanding of hate speech and counterspeech? (RQ2) How do brainstorming sessions affect users’ confidence and ability to identify key elements of hate speech and relevant counterspeech strategies? and (RQ3) How does collaborative co-writing with AI influence users’ perceptions of the co-authored responses?
We developed the interview questions to elicit participants’ reflections on each phase of the workflow. The questions were informed by prior work on counterspeech writing challenges \cite{Ping2024BehindTC, Mun2024CounterspeakersPU} and theories of human-AI collaboration~\cite{Louie2021ExpressiveCA}. 

All interviews were transcribed and analyzed using thematic analysis. Five researchers conducted open coding to identify themes related to user learning, ideation, authorship, and system usability. Codes were iteratively refined through discussion until consensus was reached. Representative quotes were selected to illustrate each major theme.

\section{Result}
\subsection{Insights from the Learning Session}

\textbf{Learning About Online Hate Speech and Its Impact.}
The learning session educated participants to better distinguish between explicit and implicit hate speech. While explicit hate speech was widely recognized, many (19; P1–P13, P15–P20) initially struggled with identifying implicit forms. As P7 noted, 

\begin{quote}
\textit{Before the session, I knew explicit hate speech was bad, but I didn’t really think this (implicit hate speech) could also be hateful.}
\end{quote}

After the session, 15 participants (P1–P5, P7, P8, P10–P13, P15–P20) expressed surprise at how subtle yet harmful implicit hate speech can be. P13, who was writing counterspeech for the first time, observed, 

\begin{quote}
\textit{After finishing the learning session, I feel that because implicit hate speech is subtle, it requires deeper reflection to identify.}
\end{quote}

P12 added that implicit hate speech is often normalized in daily discourse:

\begin{quote}
\textit{Now, I think implicit is harder to recognize because many people are not aware that what they're saying is actually hate speech. They might think it's just a normal thing to say, but there's implicit[ly] hateful meaning behind it. So, implicit hate speech can be trickier to point out because it's more subtle and socially normalized.}
\end{quote}

\textbf{Understanding Counterspeech and Its Impact.}
Before the session, participants had varied familiarity with the term ``counterspeech." Some (14; P1–P4, P8, P10–P13, P14–P17, P19) understood it as a general response to hate, while others (6; P5, P6, P7, P9, P18, P20) were less familiar. After the session, most (18; P2–P12, P14–P20) described counterspeech as a respectful and constructive way to respond to hate. P4 shared:

\begin{quote}
\textit{I did respond to some hate before, but I never knew the term. Counterspeech is a respectful way to address hate speech. It addresses the targets as human, and we should treat each other kindly. It's a way to respond to hate without more hate, but get some understanding.}
\end{quote}

\textbf{Discovering Empathy-Based Responses.}
Few participants (4; P2, P4, P8, P16) were aware of empathy-based counterspeech prior to the session. Afterward, most (19; P1–P12, P14–P20) recognized its potential to de-escalate conflict and encourage reflection.

Participants highlighted two key benefits: calming hostile exchanges (9; P2, P4, P6–P9, P13, P18, P20) and evoking empathy in the hate speaker (8; P1, P2, P4–P7, P18, P19). P6 reflected:

\begin{quote}
\textit{I liked learning about empathy-based counterspeech. If someone posted a hateful comment about gay people, I could respond like: “Hey, just so you know — my son is gay, and he's a lot more than what you said in your comment. He has dreams, hopes, and feelings.” By sharing my story, I can encourage people to reconsider their views.}
\end{quote}

Overall, our learning session expanded participants’ understanding of hate speech and counterspeech, equipping them with empathetic strategies for responding online.

\subsection{Insights from the Brainstorming Session}
\textbf{Interactive Highlighting of Hate Speech Elements.} The brainstorming session began with a color-coded highlighting exercise that educated participants identify two core components of hate speech: (1) aspects of a person or group's identity, and (2) language that dehumanizes them. Before the exercise, most participants (19; P1–P13, P15–P20) struggled to identify these components. As P11 reflected: 

\begin{quote}
\textit{I knew hate speech targeted certain groups..., but I couldn't really break it down further than that.}
\end{quote}

After completing the task, 17 participants (P1–P13, P16, P17, P19, P20) found the color-coding system helpful in making the abstract elements of hate speech more concrete. P5 explained:

\begin{quote}
\textit{The highlighting made it easy to get the main points. When I read the example, I got what was being said and the negative things that targeted the group. It showed what the hate speech was about and the specific bad things.}
\end{quote}

Many participants (15; P1–P7, P9–P12, P15, P17, P19, P20) also appreciated the system’s feedback, which included a side-by-side comparison of their highlights with the reference highlights. P13 shared:

\begin{quote}
\textit{After comparing my highlights with the system's suggestions, I see them as a good reference to better understand the hate speech. It gave me a sense of additional aspects to consider—like the focus on women, not just independent women, and how even mentioning instinct is part of the hate speech.}
\end{quote}

\textbf{Stimulating Reflective Ideation Through Guided Questions.}
Following the highlighting, the session prompted participants to answer two guided reflection questions about the hate speech: the harmful stereotypes implied, and the emotional impact on the targeted group. These questions served as a bridge from analysis to idea generation, effectively breaking down the writing task into manageable parts.

Before the exercise, participants responded to hate speech with mixed instincts—some (8; P3, P5, P8, P11, P13, P14, P18, P19) described feeling anger or frustration, while others (6; P1, P4, P6, P7, P10, P17) were unsure how to respond constructively. After engaging with the reflection prompts, many (16; P1, P2, P4, P6–P11, P13–P18, P20) noted a shift toward more thoughtful and empathetic responses. P8 described:

\begin{quote}
\textit{When I was answering the first question, I was thinking about why this was wrong and trying to explain it. In the second part, I tried to show some empathy. If I didn't answer those questions, I might have been more offensive. But now I feel like I can write it more constructively and say why it's wrong and how the targeted people might feel.}
\end{quote}

Several participants (6; P2, P7–P9, P16, P20) said the exercise prompted them to consider perspectives they had previously overlooked. P2 reflected:

\begin{quote}
\textit{In the past, I wasn't asking myself questions like this—what are the stereotypes, how does it impact the person? It's not something I thought about. This really helped me engage in that empathy-based counterspeech we talked about. It helped me see what's happening and why it's harmful.}
\end{quote}

In addition, most participants (16; P1–P7, P9–P12, P15, P17, P19, P20) found the system’s suggestions based on their responses helpful in transitioning from reflection to writing. P5 shared:

\begin{quote}
\textit{After I got the feedback, there were things I wanted to add to my input. The feedback gave me a good starting point. If you asked me to write a counterspeech from scratch, I’d start from uncertainty. But learning the feedback gave me a clear target to move on.}
\end{quote}

Twelve participants (P1–P6, P9–P12, P19, P20) emphasized that the feedback helped them think more deeply about the hate speech elements they were responding to. As P20 explained:

\begin{quote}
\textit{The feedback is good because it responds to what I actually said. It identifies useful points for writing counterspeech and expands on why those points matter. It helped me see how my own ideas could be shaped into different kinds of responses. The format and personalized nature of the feedback were really helpful.}
\end{quote}

Overall, the brainstorming session broke down the complex task of writing counterspeech into structured, guided steps—helping users analyze, reflect, and ideate in a supportive environment.

\subsection{Insights from the Co-Writing Session}
\label{s-workflow}

\textbf{Transforming Brainstormed Ideas into Counterspeech.}
\label{s-writing}

Participants found the writing interface intuitive and supportive, allowing them to focus on content without being distracted by complex features. Even participants without prior experience in AI or technical tools noted that the natural language interface made the system approachable and easy to use. As P14 reflected:

\begin{quote}
\textit{Even though I’m not familiar with AI or tech stuff, I didn’t feel lost. The interface felt natural—like I was just typing in a Google Doc—and that helped me actually focus on what I wanted to say.}
\end{quote}

Before using CounterQuill, many participants (14; P1–P6, P10, P11, P13, P14, P17–P20) expressed uncertainty about where to begin. P5 shared:

\begin{quote}
\textit{Before using CounterQuill, I wasn't sure where to start when writing counterspeech. I had ideas, but I didn't know how to structure them or make them sound convincing.}
\end{quote}

The integration of DG1 and DG2 outputs—namely the learning insights and user-created brainstorming notes—served as scaffolding for DG3. Participants (11; P2, P5, P6, P10–P12, P14, P17–P20) said that having access to their notes helped them structure their writing and include key points. P18 described:

\begin{quote}
\textit{Looking back at the notes was definitely helpful. I tried to structure my response based on the main points I had taken in the earlier sessions. Breaking things down in the notes beforehand really helped me form the counterspeech.}
\end{quote}

Several participants (14; P2–P6, P9, P11, P12, P14, P15, P17–P20) noted that starting from their own ideas reduced cognitive load and helped them feel confident in their writing. As P8 explained:

\begin{quote}
\textit{Starting a draft based on my own ideas made the writing process less stressful. I felt like I had an initial point, and it increased my confidence because I knew where to start.}
\end{quote}

\textbf{Achieving a Sense of Ownership Through Collaborative Revision}
\label{s-own}

After drafting, participants could selectively refine specific sentences with AI assistance. Most (16; P1, P3, P5–P11, P14, P15–P20) found this feature valuable for customizing tone and expression without overwriting their voice. P1 emphasized:

\begin{quote}
\textit{I think the ability to select specific sentences for rewriting is valuable. I wrote a pretty good draft and wanted to keep most of it, but there were a few sentences I wasn’t sure how to phrase. It helped that I could edit part by part while preserving the overall structure.}
\end{quote}

Participants (14; P1–P7, P12–P16, P19, P20) described a sense of ownership during writing, attributing this to their ability to retain original phrasing and make targeted revisions. P7 captured this experience:

\begin{quote}
\textit{CounterQuill's selective rewrite is helpful because I can keep what I wrote originally while also editing it part by part together with the AI. The final response feels like it's created by both of us.}
\end{quote}

Many participants (12; P2–P6, P10, P11, P12, P17–P20) said grammar checking was their first step when polishing responses. All participants used the empathetic rephrasing feature, which helped them improve clarity and relatability. Seventeen participants (P1–P8, P10–P13, P15–P18, P20) said it made their counterspeech more humanizing. As P2 described:

\begin{quote}
\textit{Using the system to polish my draft really helped. It made my comment a little shorter but also more empathetic. It kept my original meaning but added something like “This is an opportunity for us to work together.” That made it feel less confrontational and more understanding.}
\end{quote}

Participants also reported that this collaborative revision process gave them confidence. They appreciated how the system supported—not replaced—their voice, enabling a smoother transition from ideation to writing.

\section{DISCUSSION}


Our findings provide insights for designing AI-assistant tools that empower users in online counterspeech writing. We discuss three key design implications in this context.

\subsection{Structuring AI Guidance to Educate Users in Complex Counterspeech Writing}
Effective counterspeech requires careful consideration of tone, audience, and content—factors that many users find difficult to navigate without prior training \cite{Ping2024BehindTC}. Participants described feeling uncertain or overwhelmed when attempting to write unaided responses. CounterQuill’s structured, educational workflow (comprising learning, brainstorming, and co-writing) taught users how to approach counterspeech as a deliberate, multi-step process (\S~\ref{s-workflow}), making the task more approachable.

Rather than simply assisting users with content generation, CounterQuill functions as an interactive learning environment. Features such as guided highlighting, reflective questioning, and selective rewriting are designed to teach users how to identify harmful rhetoric, ideate empathetic responses, and iteratively develop their message. Our approach aligns with prior research on the benefits of structured learning environments and user agency in writing systems \cite{zhang2023visar,Caspi2011CollaborationAP}. As P5 noted, “If you asked me to write a counterspeech from scratch, I’d start from uncertainty. But learning the feedback gave me a clear target to move on.”

This educational framing was helpful for participants who lacked prior experience with AI or counterspeech writing. Several users (e.g., P14, P18), who did not have a background in computer science or experience using AI tools, described the interface as intuitive and the structured process as reducing the emotional and cognitive burden of getting started. As P18 reflected, “Breaking things down in the notes beforehand really helped me form the counterspeech.” 

While this approach may take more time than fully automated generation, it enables deeper learning, reflection, and agency~\cite{turakhia2024generating, yuan2024generative}. By guiding users to analyze hate speech, articulate their ideas, and revise their language, CounterQuill helps them develop the skills that extend beyond the immediate task. This trade-off between speed and depth means that while a multi-step workflow may not be ideal for highly experienced users seeking quick responses, it offers meaningful value for those aiming to build lasting counterspeech skills.

Future systems should continue to support both guided and flexible writing paths, adapting to users’ varying levels of confidence and experience~\cite{mouchel2023understanding,banawan2023future}, whether they benefit from scaffolded support or prefer a more autonomous writing experience.

\subsection{Fostering Perceived Authorship in Collaborative Learning in AI Co-Writing}

A central challenge in both AI-mediated writing and AI-supported learning environments is ensuring that users retain a sense of authorship over the final output~\cite{wasi2024llms,Louie2021ExpressiveCA,Pierce2003TheSO,Louie2020NoviceAIMC}. In our study, participants described their experience with CounterQuill as an opportunity to learn how to write counterspeech, supporting not only their self-expression but also their ability to reflect, organize, and revise their responses.

Participants consistently highlighted that authorship was rooted in having control over revision and expression. Many appreciated the ability to adjust specific parts of their writing while preserving their original contributions. P7 articulated this clearly:

\begin{quote}
\textit{CounterQuill's selective rewrite is helpful because I can keep what I wrote originally while also editing it part by part together with the AI. The final response feels like it's created by both of us.}
\end{quote}

This interaction reflects how AI systems can serve as collaborative learning partners, guiding users through revision and helping them develop confidence and communicative clarity. Rather than entirely producing the text on behalf of the user, CounterQuill supports a learning process in which users engage meaningfully with their own ideas.

Our findings point to the importance of writing tools that preserve user input. As prior work has shown, excessive automation may lead to disengagement and reduce users’ sense of agency \cite{lee2022coauthor,Ping2024BehindTC}. In contrast, systems designed to support transparent collaboration—such as CounterQuill’s selective rewrite and idea-tracking features—affirm users' authorship by offering guidance without erasing their voice.

Future work may further explore how interaction design supports perceptions of authorship and trust~\cite{Louie2021ExpressiveCA,Pierce2003TheSO,Louie2020NoviceAIMC}. Investigating constructs such as psychological ownership and attribution of authorship can help clarify how collaborative systems foster agency and promote authentic expression.

\subsection{Extending AI-Mediated Learning to Educational Contexts}

Our findings demonstrate that CounterQuill functions as an educational system that teaches users to engage in structured learning. Through its three-phase workflow, users participated in a sequence of activities grounded in effective educational practice: identifying harmful patterns, constructing empathetic counterspeech, and revising based on targeted feedback (§VI-A–C). These phases reflect core principles of educational design, including scaffolded support~\cite{sharma2007scaffolding}, guided reflection~\cite{canziani2013guided}, and iterative feedback~\cite{yen2017listen,nguyen2017iterative}. Prior research has shown that such structured learning system—commonly found in writing instruction and intelligent tutoring systems—can foster long-term skill development \cite{zhang2023visar,Caspi2011CollaborationAP,mouchel2023understanding,banawan2023future}.

In line with recent work on generative AI and reflective learning \cite{turakhia2024generating,yuan2024generative,banawan2023future}, participants described how specific features, such as structured reflection questions, and note-informed rewriting, taught them to “break things down,” “think more deeply,” and “rewrite with empathy” (§VI-B,C). These interactions reinforced not only the substance of counterspeech, but also the learning process of writing with empathy and rhetorical awareness.

Importantly, participants valued the ability to maintain authorship throughout the process. They emphasized that the system taught them to develop and refine their own ideas, rather than substituting them. This aligns with prior research on perceived ownership in educational co-creation systems \cite{Louie2021ExpressiveCA,10.1145/3637875,wasi2024llms}, where learning is most effective when learners remain central decision-makers~\cite{tzenios2022learner,fukuda2011facilitating,sessler2024enhancing}.

Overall, CounterQuill demonstrates that human–AI collaboration can teach users how to communicate more effectively while preserving their agency and voice. Our work offers insight that future educational systems should not only support task execution but also preserve learner agency and authorship throughout the AI-mediated learning process.

\section{CONCLUSION}
In this work, we introduced \textit{CounterQuill} to educate and support users in writing counterspeech. Our system guides users through three sequential phases: a learning session to build foundational knowledge about hate speech and counterspeech, a brainstorming session to help identify harmful language and generate response strategies, and a co-writing session that enables collaborative refinement of counterspeech with AI assistance.

Our user study showed that this structured workflow fosters both learning and engagement. The guided brainstorming process gave users confidence in analyzing hate speech, while the co-writing phase strengthened their sense of authorship and agency.

\textit{\textbf{Limitations and future work.}} A limitation of our study lies in the relatively small sample size and its reliance on qualitative feedback from interviews. While these insights offer valuable depth, the limited number of participants restricts the generalizability of our findings. In addition, we did not assess the objective quality of the counterspeech generated. Evaluating the quality and effectiveness of counterspeech requires a larger pool of data and must account for the actual reactions of hate speakers to provide a more reliable measure \cite{Ping2024BehindTC}. Our current system is designed to lay the groundwork for the future development of a long-term web-based plugin. To address these limitations, future studies should involve larger and more diverse participant samples, reaching scales in the thousands, and incorporate external evaluations such as expert reviews or crowd-sourced ratings.

Additionally, our evaluation focused on a single hate speech scenario, which may not represent the full range of online discourse. Future research should include multiple hate speech types to evaluate the system’s adaptability and generalizability.

To extend this work, we plan to develop CounterQuill as a lightweight browser extension (e.g., for Chrome and Edge) that supports co-writing with AI and provides on-demand guidance. This extension will integrate into platforms like Reddit and X, enabling users to engage with the tool in context and at their own pace. A longitudinal field deployment will further assess the system’s usability and effectiveness in real-world social media environments, and future evaluations will also incorporate direct assessments of counterspeech quality. Additionally, we plan to explore mechanisms such as fact-checking aids and community flagging to mitigate risks of misuse and AI-generated hallucinations in sensitive contexts.

\bibliographystyle{IEEEtran} 
\bibliography{citation}
\end{document}